\documentclass[aps,prb,longbibliography,reprint]{revtex4-1}

\usepackage{graphicx}
\usepackage{amsmath}
\usepackage{amsfonts}
\usepackage{amssymb}
\usepackage{bbm}
\usepackage[]{units}
\usepackage{hyperref}
\usepackage{braket}

\newcommand{\nnb}{\nonumber \\}

\newcommand{\bv}{\left( \begin{array}{c}}
\newcommand{\ev}{\end{array} \right)}
\newcommand{\E}{\mathrm{e}}

\newcommand{\st}[1]{_{\text{#1}}}

\newcommand{\I}{\mathrm{i}}

\newcommand{\braketv}[2]{\left\langle #1\right|\left.\! #2 \right\rangle }
\begin{document}
\title{Coupling of three-spin qubits to their electric environment}
\author{Maximilian Russ, Florian Ginzel, and Guido Burkard}
\affiliation{Department of Physics, University of Konstanz, D-78457 Konstanz, Germany}
\begin{abstract}
We investigate the behavior of qubits consisting of three electron spins in double and triple quantum dots subject to external electric fields. Our model includes two independent bias parameters, $\varepsilon$ and $\varepsilon_{M}$, which both couple to external electromagnetic fields and can be controlled by gate voltages applied to the quantum dot structures. By varying these parameters one can switch the qubit type by shifting the energies in the single quantum dots thus changing the electron occupancy in each dot. Starting from the asymmetric resonant (ARX) exchange qubit with a (2,0,1) and (1,0,2) charge admixture one can smoothly cross over to the resonant exchange (RX) qubit with a detuned (1,1,1) charge configuration, and to the exchange-only (EO) qubit with the same charge configuration but equal energy levels down to the hybrid qubits with (1,2,0) and (0,2,1) charge configurations. Here, ($l,m,n$) describes a configuration with $l$ electrons in the left dot, $m$ electrons in the center dot, and $n$ electrons in the right dot.
We first focus on random electromagnetic field fluctuations, i.e., ``charge noise'', at each quantum dot resulting in dephasing of the qubit and provide a complete map of the resulting dephasing time as a function of the bias parameters. We pay special attention to the so-called sweet spots and double sweet spots of the system which are least susceptible to noise. 
In the second part we investigate the coupling of the qubit system to the coherent quantized electromagnetic field in a superconducting strip-line cavity and also provide a complete map of the coupling strength as a function of the bias parameters. We analyze the asymmetric qubit-cavity coupling via $\varepsilon$ and the symmetric coupling via $\varepsilon_{M}$.
\end{abstract}
\maketitle
%%%%%%%%%%%%%%%%%%%%%%%%%%%%%%%%%%%%%%%%
\section{Introduction}

Qubits based on the spin of electrons trapped in quantum dots (QDs)\cite{Kloeffel2013} are a leading candidate for enabling quantum information processing. They provide long coherence times\cite{Petta2005,Greilich2006,Koppens2008,Morello2010,Maune2012,Pla2012,Yang2013,Awschalom2013,Muhonen2014,Eng2015,Reed2016}, together with a scaleable architecture for a dense qubit implementation. Semiconductor materials like gallium arsenide (GaAs)\cite{Hanson2007} and silicon\cite{Zwanenburg2013} are the most common choices as the host material for QDs. One common feature of these implementations is the need for control with electric fields at the nanoscale which unavoidably couples the qubit system to electrical noise\cite{Kloeffel2013}. Dominating sources of decoherence are nuclear spins\cite{Coish2004,Mehl2013,Hung2014}, spin-orbit interaction\cite{Glazov2004,Prada2016}, and charge noise from either the environment or the confining gates\cite{Yurkevich2010,Dial2013,Taylor2013,Fei2015,Russ2015,Shim2016}. The effect of the first and second source of decoherence can be drastically reduced by using silicon as the host material due to its highly abundant nuclear spin free isotope and a weak spin-orbit interaction\cite{Dial2013}. Using active noise suppression methods such as quantum error correction\cite{Lidar2013} and composite pulse sequences\cite{West2012,Setiawan2014,Rohling2015} leaves charge noise coupled to the spin as the remaining problem to be taken care of. Thus, additional passive suppression methods are needed such as optimal working points (sweet spots)\cite{Vion2002,Paladino2014} which vary in effectiveness for different qubit implementations. 

\begin{figure}[t]
\begin{center}
\includegraphics[width=1.0\columnwidth]{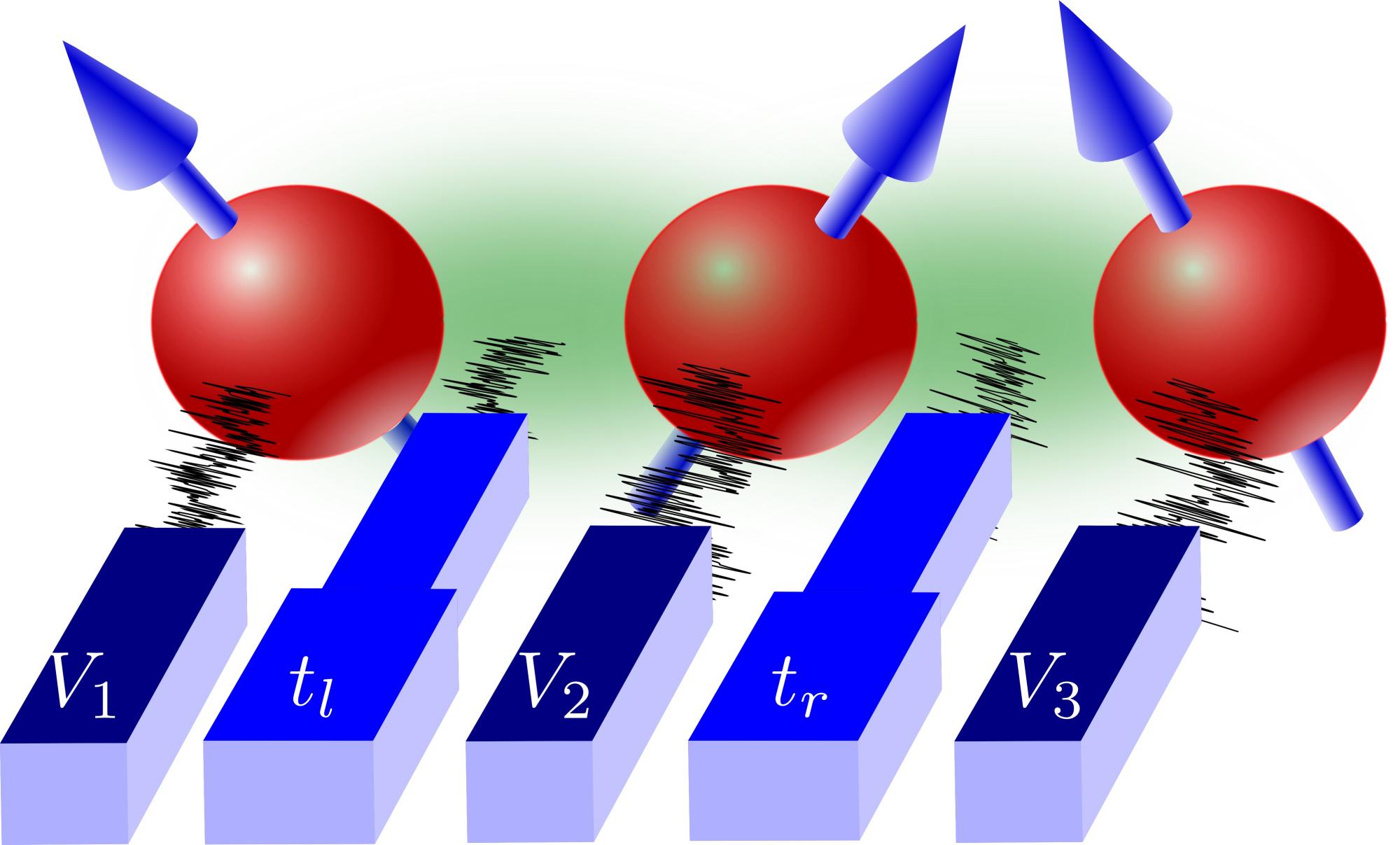}
\caption{Schematic illustration of a three-spin qubit coupled to a noisy electric environment. The environment can affect the electron spins directly through the gate voltages $V_{i}$ with $i\in\lbrace 1,2,3\rbrace$ of each quantum dot (QD) or the exchange coupling (green cloud) between the electron spins through the gate-controlled tunnel hopping ($t_{l}$ and $t_{r}$).}
\label{fig:model}
\end{center}
\end{figure}

Qubit implementations using single or multiple QDs to encode a single qubit show high-fidelity gate operations, long decoherence times together with fast qubit control allowing for many operations during the qubit lifetime\cite{Takakura2010,Gaudreau2012,Maune2012,Medford2013,Veldhorst2015}. An advantage of multi-spin qubit encodings consists in their improved protection against certain types of noise\cite{Lidar2003} together with faster gate operations\cite{nature2000,Laird2010,Shi2012,Koh2012,Cao2016}. This ultimately leads to the three-spin-$\frac{1}{2}$ qubits (see Fig.~\ref{fig:model}); the exchange-only (EO) qubit allowing for full qubit control with only the exchange interaction\cite{nature2000}, the resonant exchange (RX) qubit with permanently acting exchange interaction and control through resonant driving\cite{Medford2013,Taylor2013,Russ2015}, and the always-on exchange-only (AEON) qubit with symmetric gate control\cite{Shim2016}. Robustness against charge noise can be achieved by operating the qubit on sweet spots\cite{Taylor2013} where the qubit energy splitting is extremal with respect to one noisy parameter or double sweet spots\cite{Russ2015b,Shim2016} where both noisy parameters are optimized. In this paper we provide a full analysis of charge noise for three-spin-$\frac{1}{2}$ qubits (Fig.~\ref{fig:model}) and present optimal working points. We go beyond previous work\cite{Russ2015,Shim2016} by exploring the full $(\varepsilon,\varepsilon_{M})$ parameter space. Moreover, we include two noisy tunneling parameters mapping the resulting dephasing time in this parameter space. Single sweet spots (SSS) and double sweet spots (DSS) are presented for both types of noisy parameters and combined to provide the best working points.

Two-qubit gates are provided by the exchange interaction\cite{nature2000,Fong2011,Setiawan2014,Doherty2013,Wardrop2016,Shim2016}, Coulomb interaction\cite{Taylor2013,Pal2014,Pal2015} and cavity-mediated coupling\cite{Russ2015b,Srinivasa2016} while the range of the latter is only limited by the extension of the cavity. This long-ranged coupling technique can be realized within the approach of cavity electrodynamics (cQED) by coupling the qubit capacitely to a superconducting strip-line cavity\cite{Blais2004,Wallraff2004}  adapted for spin qubits\cite{Imamoglu1999,Childress2004,Burkard2006,Taylor2006,Hu2012,Petersson2012,Russ2015b,Srinivasa2016,Beaudoin2016a,Royer2016,Beaudoin2016b}. Implementing a three-spin qubit in a triple quantum dot (TQD) coupled to a cavity is possible for two distinct setups; a longitudinal coupling or asymmetric setup and a transversal or symmetric coupling\cite{Russ2015b}. In this paper both of these setups are discussed, going beyond previous work for the asymmetric implementation\cite{Russ2015b} and providing a microscopic description for both implementations.

This paper is organized as follows. In section~\ref{sec:qubit}, we define the three-spin qubit states, discuss the different regimes in parameter space where each qubit implementation is located and their conversion into each other. In section~\ref{Sec:chargedephasing}, we analyze in detail the optimal working points best suitable for operating the qubit in the presence of charge noise coupled to the qubit through detuning and tunnel parameters. Subsequently in section~\ref{sec:cqed}, we present two setups for coupling three-spin qubits to a superconducting strip-line cavity in order to find operation points with a strong and controllable coupling. We conclude in section~\ref{sec:conc} with a summery and outlook.

\section{Qubit}
\label{sec:qubit}
We consider the spins of three electrons in a linearly arranged triple quantum dot (TQD) (Fig.~\ref{fig:model}) where each QD has a single available orbital, whereas, higher orbitals are energetically unfavorable due to a strong confinement. Additionally, we restrict ourselves to the spin degree of freedom (DOF) only, hence, we either consider a material with no valley DOF or a strong valley splitting surpassing the energy of the exchange splittings and then treat the valley as an orbital DOF. Further we assume that the TQD is connected to an electric environment (schematically illustrated in Fig.~\ref{fig:model}) via the gate voltages $V_{i}$ of QD~$i$ with $i\in\lbrace 1,2,3\rbrace$ and via the gate-controlled tunnel barriers, consisting of either random electric fluctuations or a coherent quantized electric field. As a model of the TQD we use the three-site extended Hubbard model
\begin{align}
H\st{Hub}=&\sum_{i=1}^{3} \left[\frac{\tilde{U}}{2}n_i(n_i-1) + V_in_i\right] \nnb
&+ \sum_{\braket{i,j}}\left[U_{C}n_in_j + \sum_{\sigma=\uparrow,\downarrow}\left(t_{ij}c_{i,\sigma}^\dagger c_{j,\sigma}+\text{h.c.}\right)\right],
	\label{eq:Hub}
\end{align}
where $c_{i,\sigma}^{\dagger}$ ($c_{i,\sigma}$) creates (annihilates) an electron in QD $i$ with spin $\sigma$. We define the number operator $n_i=\sum_\sigma c_{i,\sigma}^\dagger c_{i,\sigma}$ and the gate-controlled pairwise hopping matrix elements $t_{ij}$ with $i,j\in \{1,2,3\}$. Here, we consider symmetric, spin-conserving nearest-neighbor hopping, $t_{13}=t_{31}=0$, $ t_{12}=t_{21}\equiv t_{l}/ \sqrt{2}$, and $ t_{23}=t_{32}\equiv t_{r}/ \sqrt{2}$. We also include the Coulomb repulsion of two electrons in the same QD $\tilde{U}$ and in neighboring QDs $U_{C}$ which leads to a static energy shift in the dots. 

\begin{figure}
\begin{center}
	\includegraphics[width=1.0\columnwidth]{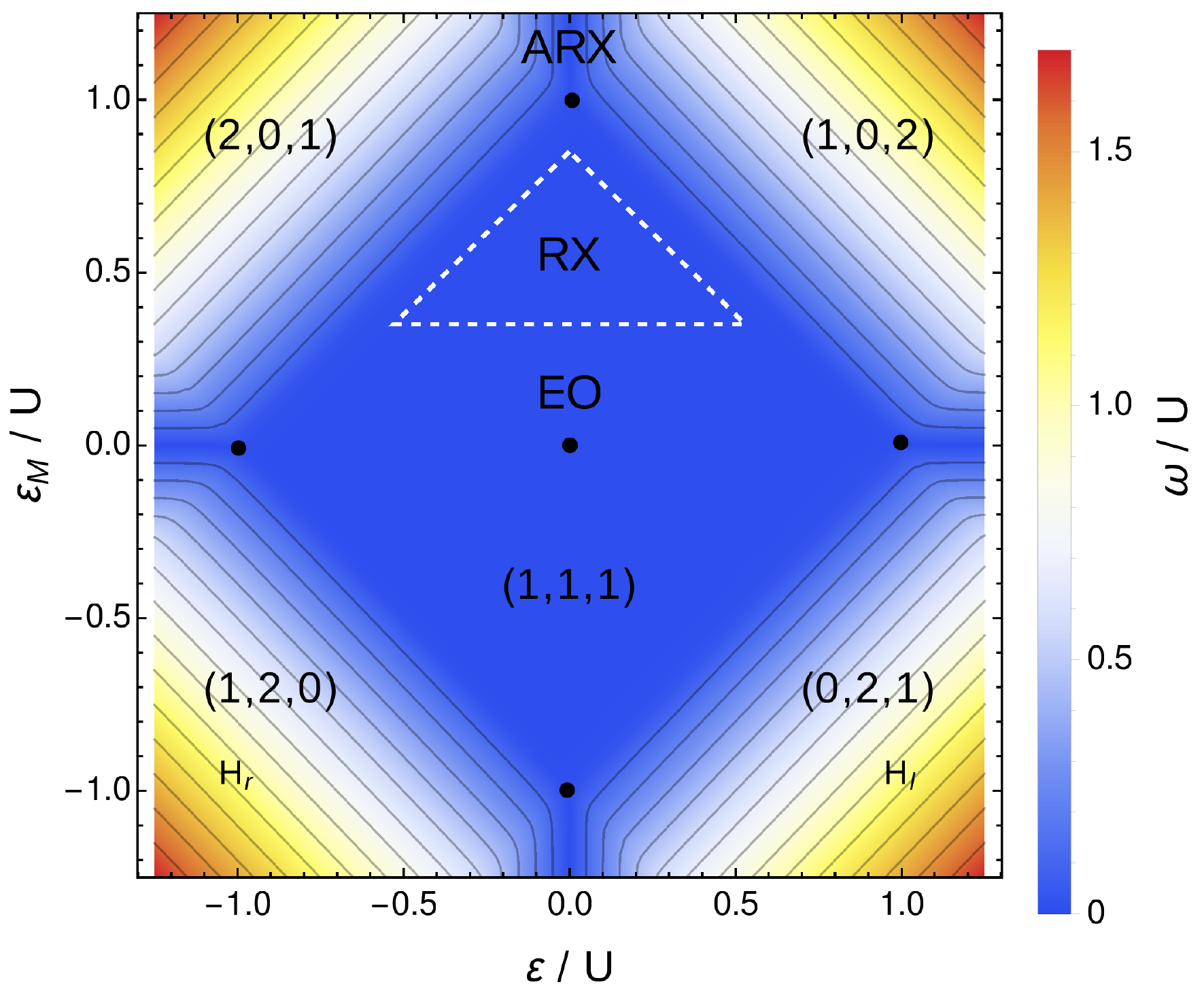}
	\caption{Energy landscape of the ground-state energy gap $\omega$ as a function of the detuning parameters $\varepsilon$ and $\varepsilon_{M} $ in units of the on-site Coulomb repulsion $U$. Maneuvering through the ($\varepsilon$, $\varepsilon_{M} $)~plane one can access various parameter regimes that allow the use of different qubit implementations in different charge configurations $(l,m,n)$, where $l$ electrons are in the left, $m$ electrons in the center, and $n$ electrons in the right QD. We further highlight the double sweet spots (DSS) (black dots), the location of the exchange-only (EO) qubit, the resonant exchange (RX) qubit (dashed triangle), the asymmetric resonant exchange (ARX) qubit, and the left and right hybrid (H$_{l,r}$) qubit.}
	\label{fig:landscape}
\end{center}
\end{figure}

Restricting ourselves to the subspace of the three-spin Hilbert space with total spin $S=S_{z}=1/2$, we identify six relevant states, two states with a (1,1,1) charge configuration, and one each with a (2,0,1), (1,0,2), (1,2,0), (0,2,1) charge configuration,
\begin{align}
	\ket{0}&\equiv \frac{1}{\sqrt{2}}\left(c_{1,\uparrow}^\dagger c_{2,\uparrow}^\dagger c_{3,\downarrow}^\dagger-c_{1,\downarrow}^\dagger c_{2,\uparrow}^\dagger c_{3,\uparrow}^\dagger\right)\ket{\text{vac}}   \nnb
	&=\ket{s}_{13}\ket{\uparrow}_{2},\\
	\label{state:0}
	\ket{1}&\equiv \frac{1}{\sqrt{6}}\!\left(2c_{1,\uparrow}^\dagger c_{2,\downarrow}^\dagger c_{3,\uparrow}^\dagger\!-\!c_{1,\uparrow}^\dagger c_{2,\uparrow}^\dagger c_{3,\downarrow}^\dagger\!-\!c_{1,\downarrow}^\dagger c_{2,\uparrow}^\dagger c_{3,\uparrow}^\dagger\right)\!\ket{\text{vac}},  \nnb
	&=\sqrt{\frac{2}{3}}\ket{t_{+}}_{13}\ket{\downarrow}_{2}-\frac{1}{\sqrt{3}}\ket{t_{0}}_{13}\ket{\uparrow}_{2},\\
	\label{state:1}
	\ket{2}&\equiv  c_{1,\uparrow}^\dagger c_{1,\downarrow}^\dagger c_{3,\uparrow}^\dagger\ket{\text{vac}}=\ket{s}_{11}\ket{\uparrow}_{3},   \\
	\ket{3}&\equiv c_{1,\uparrow}^\dagger c_{3,\uparrow}^\dagger c_{3,\downarrow}^\dagger\ket{\text{vac}}=\ket{\uparrow}_{1}\ket{s}_{33},\\
	\ket{4}&\equiv c_{1,\uparrow}^\dagger c_{2,\uparrow}^\dagger c_{2,\downarrow}^\dagger \ket{\text{vac}}=\ket{\uparrow}_{1}\ket{s}_{22},   \\ 
	\ket{5}&\equiv c_{2,\uparrow}^\dagger c_{2,\downarrow}^\dagger c_{3,\uparrow}^\dagger\ket{\text{vac}}=\ket{s}_{22}\ket{\uparrow}_{3},
	\label{eq:logicalspace}
\end{align}
where $ \ket{\text{vac}} $ denotes the vacuum state, $\ket{s}_{ij}$ denotes the singlet state, $\ket{t_{0}}_{ij}$ denotes the $S_{z}=0$ triplet state, and $\ket{t_{+}}_{ij}$ denotes the $S_{z}=+1$ triplet state. Here, ($l,m,n$) describes a configuration with $l$ electrons in the left dot, $m$ electrons in the center dot, and $n$ electrons in the right dot. An additional leakage state with $S=3/2$ and $S_{z}=1/2$ is not coupled since charge noise conserves the total spin $S$.

We introduce a new set of detuning parameters $V\st{tot}=(V_{1}+V_{2}+V_{3})/3$, $\varepsilon\equiv (V_{1}-V_{3})/2$ and $\varepsilon_{M}\equiv V_{2} - (V_{1}+V_{3})/2+U_{C}$. The parameter $V\st{tot}$ merely shifts the total energy, hence, contributes to neither the dynamics of the qubit nor the decoherence. The asymmetric detuning $\varepsilon$ is the energy difference between the left QD and the right QD, and the symmetric detuning $\varepsilon_{M}$ is the energy difference between the center QD and the mean of the outer QDs. Defining the charging energy $U\equiv \tilde{U}-U_{C}$ we find for the Hamiltonian Eq.~\eqref{eq:Hub} in the $S=S_{z}=1/2$ basis $\left\lbrace \ket{0},\ket{1},\ket{2},\ket{3},\ket{4},\ket{5}\right\rbrace$ the matrix representation
\begin{align}
H=\left(
	\begin{smallmatrix}
	 0 & 0 & \frac{1}{2}t_{l} & \frac{1}{2}t_{r} & \frac{1}{2}t_{r}  & \frac{1}{2}t_{l}  \\
	 0 & 0 & \frac{\sqrt{3}}{2} t_{l} & -\frac{\sqrt{3}}{2} t_{r} & -\frac{\sqrt{3}}{2} t_{r} & \frac{\sqrt{3}}{2} t_{l} \\
	 \frac{1}{2}t_{l} &  \frac{\sqrt{3}}{2}t_{l} & \varepsilon -\varepsilon_{M}+ U& 0 & 0 & 0 \\
	 \frac{1}{2}t_{r} & -\frac{\sqrt{3}}{2} t_{r} & 0 & -\varepsilon -\varepsilon_{M}+ U & 0 & 0\\
	 \frac{1}{2}t_{r} & -\frac{\sqrt{3}}{2} t_{r}& 0 & 0  & \varepsilon +\varepsilon_{M}+ U & 0 \\
	 \frac{1}{2}t_{l} & \frac{\sqrt{3}}{2} t_{l}  & 0 & 0 & 0 & -\varepsilon +\varepsilon_{M}+ U\\
	\end{smallmatrix}
	\right).
	\label{eq:hubmatrix}
\end{align}
We assume $t_{l,r}$ to be real since any complex phase only contributes a global phase to the eigenstates.

Depending on the position in the ($\varepsilon$, $\varepsilon_{M} $)-plane different qubit implementations are realized (Fig.~\ref{fig:landscape}). Directly in the center of the (1,1,1) charge occupancy region, the \textit{exchange-only (EO) qubit}\cite{nature2000} and the always-on exchange-only (AEON) qubit\cite{Shim2016} are located. Still in the (1,1,1) charge occupancy, but in the area with $\varepsilon_{M}\gg\varepsilon$ which allows transitions into the (2,0,1) and (1,0,2) charge states, we find the \textit{resonant exchange (RX) qubit} (white dashed triangle). The \textit{asymmetric resonant exchange (ARX) qubit} is located deeper into the regime with $\varepsilon_{M}\gg\varepsilon$ and a strong mixture of (2,0,1) and (1,0,2) charge configurations\cite{Russ2015}. Due to mirror-symmetry four DSS can be found (black dots) in the corner of the diamond-shaped (1,1,1)-charge configuration area. At the bottom left (right) in the ($\varepsilon$, $\varepsilon_{M} $)-plane in Fig.~\ref{fig:landscape} only two neighboring QDs are occupied by three electrons giving rise to the double quantum dot (DQD) \textit{hybrid qubit}\cite{Shi2012,Koh2012,Cao2016} formed in QD~1 and QD~2 (QD~2 and QD~3).

\section{Charge dephasing}
\label{Sec:chargedephasing}

Recent progress of spin qubits using purified silicon as the host material show exceptionally long $T_{1}$ and $T_{2}$ times emphasizing the importance of charge noise. The use of isotopically purified Si eliminates nuclear spins as the leading source for decoherence and leaves charge noise as the main cause of decoherence\cite{Dial2013}. Charge noise or electrical noise originates from random charge fluctuations of the material or from the control and confinement voltages giving rise to fluctuating energies. Formally, we describe this by substituting $q\rightarrow q + \delta q$ in which the parameter $q\in\lbrace\varepsilon,\varepsilon_{M}, t_{l}, t_{r}\rbrace$ is affected by random fluctuations $\delta q\in\lbrace\delta \varepsilon,\delta \varepsilon_{M},\delta  t_{l},\delta  t_{r}\rbrace$. There are two effects of decoherence for charge noise, longitudinal and transversal dephasing, where the first causes the energy gap between the qubit states $\ket{e}$ and $\ket{g}$ to fluctuate while the second one gives rise to transitions between the qubit states. These can further be divided into decoherence due to detuning parameters\cite{Taylor2013,Fei2015,Russ2015,Shim2016} ($\varepsilon$, $\varepsilon_{M}$) and decoherence due to charge noise coupled to the qubit by tunneling ($t_{l}$, $t_{r}$).

The remainder of this section is organized as follows. First we present a generalized framework extending previous models describing charge noise coupled to the qubit to different control parameters. In the next step, we take only individual control parameters into account. We start with longitudinal and transversal noise originating from the detuning parameters $\varepsilon$ and $\varepsilon_{M}$ and identify sweet spots\cite{Taylor2013,Fei2015,Russ2015,Shim2016}, i.e., working points in which the qubits lifetime is highly increased due to vanishing coupling in first order of the qubit to the noise. Subsequently, we focus on the effects of noise coupled to the qubit via fluctuations of the tunnel amplitude and show that there exist no sweet spots for both noisy tunneling parameters simultaneously. In the last part of this section, we take all separately discussed effects of the noisy parameters into account in order to present the best working points.

\subsection{General framework}
\label{ssec:genframe}
In general, we can write the qubit Hamiltonian in its eigenbasis\cite{Russ2015}, $\lbrace\ket{g},\ket{e}\rbrace$, as
\begin{align}
H\st{noise}=\frac{\hbar}{2}\left[\left(\omega + \delta\omega_{z}\right)\sigma_{z} + \delta\omega_{x}\sigma_{x} + \delta\omega_{y}\sigma_{y}\right],
\label{eq:Hnoise}
\end{align}
with the unperturbed energy gap $\hbar\omega$ between the qubit states. Here, $\ket{e}$ and $\ket{g}$ are the two lowest eigenstates of $H$ in Eq.~\eqref{eq:hubmatrix}. The longitudinal effect of the charge noise up to second order is
\begin{align}
\delta\omega_{z}=\sum_{q}\left( \omega_{q} \delta q+\frac{\omega_{q,q}}{2}\delta q^{2}\right)+\frac{1}{2}\sum_{p\neq q} \omega_{p,q}  \delta p\delta q,
\label{eq:noiseGenz}
\end{align}
where we used the definitions $\omega_{q}\equiv\frac{\partial \omega}{\partial q}$ and $\left.\omega_{p,q}\equiv\frac{\partial^{2}\omega}{\partial p\partial q}\right.$ with $p,q\in \lbrace\varepsilon,\varepsilon_{M},t_{l},t_{r}\rbrace$. For the transversal effects we consider
\begin{align}
\delta\omega_{x}=\sum_{q} \delta\omega_{x,q} \delta q,
\end{align}
with $\delta\omega_{x,q}\equiv\bra{g} H_{1,q} \ket{e}$ and $\delta\omega_{y}=0$ due to real valued tunneling. Here, $H_{1,q}$ is the part of the Hamiltonian from Eq.~\eqref{eq:hubmatrix} associated with the perturbation in $q$, thus, $H_{1,q}=\frac{\partial}{\partial q}H\,\delta q$.
Considering only longitudinal noise one can calculate the pure dephasing in a Ramsey free decay approach\cite{Taylor2006}
\begin{align}
	\tilde{f}(t)\equiv\left\langle \E^{\I \phi(t)} \right\rangle = \E^{-\left\langle\phi(t)^{2}\right\rangle}=\E^{-t^{2}/T_{\varphi}^{2}}
	\label{eq:ramsey}
\end{align}
with $\phi(t)\equiv\int_{t_{0}}^{t} dt^{\prime} \delta\omega_{z}(t^{\prime})$ in which we used $\delta q=\delta q(t)$. For the first equality we used Gaussian distributed noise with zero mean, while for the second equality we used a spectral density exponent\cite{Freeman2016} $\gamma=1$. This allows us to calculate for a given spectral density of the noise $S_{q}(\tilde{\omega})=A_{q}|\tilde{\omega}|^{-\gamma}$ the associated dephasing time\cite{Russ2015}
\begin{align}
	T_\varphi=& \hbar \bigg[\sum_{q}\frac{\omega_{q}^2}{2}\,A_p\log r+\frac{\omega_{q,q}^2}{4}A_q^2\log^2r\nnb
	&+\frac{1}{2}\sum_{p\neq q}\frac{ \omega_{p,q}^2}{2}A_{p}A_{q}\log^2r+ \frac{1}{8}\omega_{p,p} \omega_{q,q}A_p\,A_q\bigg]^{-\frac{1}{2}}.
\label{eq:dephasing}
\end{align}
Here, we assumed independent and uncorrelated noise for each noisy parameter $p,q\in \lbrace\varepsilon,\varepsilon_{M},t_{l},t_{r}\rbrace$ and used the ratio $r\equiv\omega\st{UV}/\omega\st{IR}$ of lower frequency cutoff $\omega\st{IR}$ and upper frequency cutoff $\omega\st{UV}$ needed to ensure convergence of the intergral.

With this in mind, we formally define the expression ``sweet spot'' initially introduced as best points for operations due to a vanishing coupling of the qubit to the noise in first order. Assuming first order noise effects to be the dominant ones these points yield the longest life-time of the qubit according to Eq.~\eqref{eq:dephasing}, and therefore, the ideal operation points for the qubit. Taking only first order effects into account we obtain for the best working points the condition
\begin{align}
\sum_{q} \omega_{q} \delta q = 0,
\label{eq:SScond}
\end{align}
which is in general only possible for each $\omega_{q}=0$ with $q\in \lbrace\varepsilon,\varepsilon_{M},t_{l},t_{r}\rbrace$. We now define a single ``sweet spot'' (SSS) if this condition is fulfilled for one noisy parameter. Analogously, we define a ``double sweet spot'' (DSS) if $\omega_{q}=\omega_{p}=0$ with $q\neq p\in \lbrace\varepsilon,\varepsilon_{M},t_{l},t_{r}\rbrace$. Considering a total of four noisy parameters we can also introduce ``triple'' and ``quadruple sweet spots'' in which Eq.~\eqref{eq:SScond} is for three noisy parameters or completely fulfilled. Unfortunately, we find that there exist no such quadruple sweet spots in a three-spin-$\frac{1}{2}$ qubit in maximally three QDs.

The second term in the Hamiltonian (see Eq.~\eqref{eq:Hnoise}) $\delta\omega_{x}$ leads to random rotations of the qubit around the $x$-axis with timescales on the order of ms for realistic charge noise (assuming $\sqrt{\langle\delta q^{2}\rangle}=\unit[1]{\mu eV}$ at $\unit[1]{Hz}$)\cite{Freeman2016}. Somewhat more devastating for the qubit is the fact that transversal charge noise changes the orientation of the eigenstates and therefore the energy gap giving rise to an additional term for the dephasing in second order of the fluctuations. This becomes clear when expanding the eigenenergy difference from Eq.~(\ref{eq:Hnoise}),
\begin{align}
\omega &= \sqrt{(\omega_0+\delta\omega_z)^2+\delta\omega_x^2+\delta\omega_y^2}\nonumber\\
&\simeq  \omega_0 + \delta\omega_z + \frac{\delta\omega_x^2}{2\omega_0} + \frac{\delta\omega_y^2}{2\omega_0}
+O(\delta\omega^3).
\end{align}

\begin{figure*}
\begin{center}
\includegraphics[width=1.\textwidth]{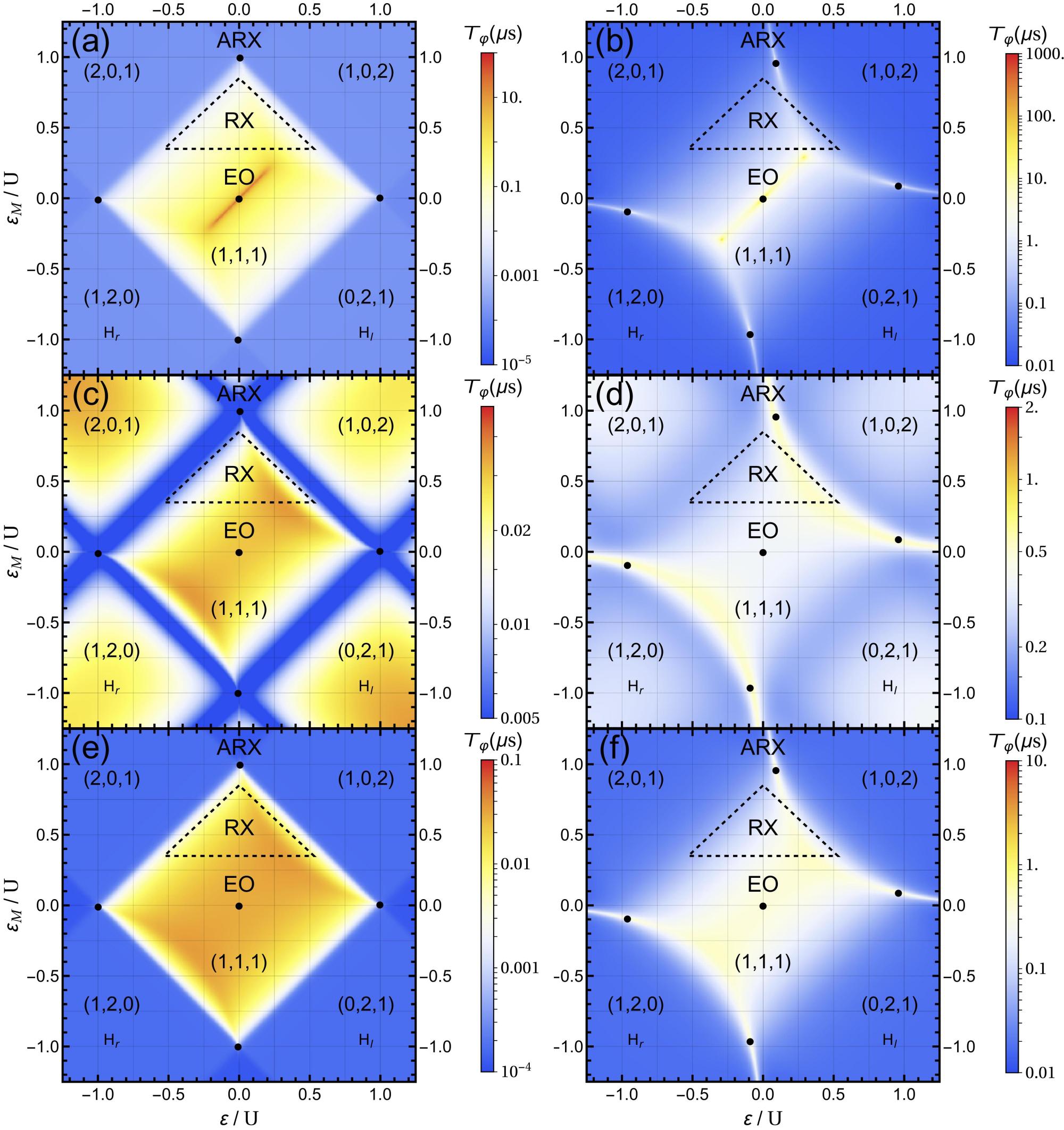}
\caption{Dephasing time $T_{\varphi}$ given by Eq.~\eqref{eq:dephasing} due to longitudinal noise as a function of the detuning parameters $\varepsilon$ and $\varepsilon_{M}$. In the top row ((a) and (b)) we plot $T_{\varphi}$ resulting from charge noise in the two detuning parameters $\varepsilon$ and $\varepsilon_{M}$, in the center row ((c) and (d)) we plot $T_{\varphi}$ resulting from charge noise in the two tunneling parameter $t_{l}$ and $t_{r}$, and in the bottom row ((e) and (f)) we plot $T_{\varphi}$ resulting from charge noise from all four parameters combined, where we choose the parameter settings identical in each column. The left column shows results for weak tunneling and strong noise while in the right column, results for strong tunneling and weak noise are plotted. Parameters are set as follows; $t_{l}=\unit[22]{\mu eV}$, $t_{r}=\unit[15]{\mu eV}$, $A_{q}=\unit[1]{\mu eV}^{2}$ where $q=\varepsilon,\varepsilon_{M}$ in (a) and (e), and $A_{q}= (\unit[10^{-1}]{\mu eV})^{2}$, where $q=t_{l},t_{r}$ in (c) and (e), for the left column and $t_{l}=\unit[220]{\mu eV}$, $t_{r}=\unit[150]{\mu eV}$, $A_{q}=(\unit[10^{-2}]{\mu eV})^{2}$ where $q=\varepsilon,\varepsilon_{M}$ in (b) and (f), and $A_{t_{l}}=A_{t_{r}}= (\unit[10^{-3}]{\mu eV})^{2}$, where $q=t_{l},t_{r}$ in (d) and (f), for the right column. To include a large frequency bandwidth we globally set the ratio of the lower and higher frequency cut-off $r=5\times10^{6}$. The black dots indicate the DSS.}
\label{fig:Plotlong}
\end{center}
\end{figure*}

Inside the (1,1,1) charge configuration regime and away from the charge transition lines, the states $\ket{0}$ and $\ket{1}$ defined in Eq.~\eqref{state:0} and Eq.~\eqref{state:1} are nearly the qubit states while the coupling of the other states can be taken into account using a low-energy Schrieffer-Wolff approximation\cite{Bravyi2011}. We obtain for the resonance frequency between the qubit eigenstates\cite{Shim2016}
\begin{align}
\hbar\omega = \sqrt{J^{2}+3j^{2}},
\label{eq:HeigenSW}
\end{align}
with the mean $J\equiv (J_{l}+J_{r})/2$ and the half of the difference $j\equiv(J_{l}-J_{r})/2$ of the left and right exchange interaction and respectively between the left (right) QD and the center~QD,
\begin{align}
J_{l}=& 2t_{l}^{2}U/\left[U^{2}-(\varepsilon-\varepsilon_{M})^{2}\right],\\
J_{r}=&2t_{r}^{2}U/\left[U^{2}-(\varepsilon+\varepsilon_{M})^{2}\right].
\end{align}
Utilizing this, we find a closed analytical expression for the longitudinal fluctuation
\begin{align}
	\omega_{q}=(J\partial_{q}J+3j\partial_{q} j)/\omega,
\end{align}
and the transversal effect
\begin{align}
	\delta\omega_{x,q}=(J\partial_{q}j-j\partial_{q}J)/\omega,
\end{align}
in the (1,1,1) charge regime. For $\varepsilon_{M}\gg 0$ (RX regime) these results converge with the expressions considering only the RX qubit\cite{Russ2015}, since there the influence of the states $\ket{4}$ and $\ket{5}$ becomes negligible. 

\subsection{Detuning noise}
\label{ssec:long}

Longitudinal dephasing $T_{\varphi}$ due to low frequency charge noise originating from the detuning parameters $\varepsilon$ and $\varepsilon_{M}$ is usually seen as the dominant source for decoherence. For reasons of simplicity we set in this subsection $\delta t_{l}=\delta t_{r}=0$ and only consider charge noise originating from the detuning parameters $ \delta\varepsilon\neq 0$ and $ \delta\varepsilon_{M}\neq 0$. The effect of this can be drastically reduced by working on SSS\cite{Taylor2013} or DSS\cite{Fei2015,Russ2015,Shim2016}. They fulfill the condition $\omega_{\varepsilon}=\omega_{\varepsilon_{M}}=0$ such that the longitudinal coupling given in Eq.~\eqref{eq:noiseGenz} vanishes and only second order effects remain. 

In Figs.~\ref{fig:Plotlong}~(a)-(b), we plot the resulting dephasing time $T_{\varphi}$ given in Eq.~\eqref{eq:dephasing} as a function of the detuning parameters considering only longitudinal noise originating from $\varepsilon$ and $\varepsilon_{M}$ for different parameter settings. We find in total five DSS for the three-spin qubit marked as black dots. Two DSS are already known, one inside the (1,1,1) charge configuration regime\cite{Shim2016} and one at the transition between the (2,0,1) and (1,0,2) charge states\cite{Russ2015}, while the other three DSS appear at the remaining three charge transitions, located on the left between (2,0,1) and (1,2,0), on the right right between (1,0,2) and (0,2,1), and bottom between (1,2,0) and (0,2,1) in the figures due to symmetry considerations. For symmetric tunneling ($t_{l}=t_{r}=t$) the five DSS are approximately located at $(\varepsilon,\varepsilon_{M})=(0,U),\,(0,-U),\,(-U,0),\,(U,0)$, and $(0,0)$, while for asymmetric tunneling ($t_{l}\neq t_{r}$) all DSS except the center one are slightly shifted due to a shift of the charge transitions. Comparing these DSS with each other, the four DSS located at the charge transitions are unfavorable for a small tunneling and strong noise due to strong higher order effects limiting the coherence of the qubit. Considering stronger tunneling between the QDs and weaker noise, the higher order effects are greatly reduced due to softening of the charge transitions. If for some reasons working on the DSS is unpractical, e.g. coupling the qubit to a cavity, one should favor in the case $t_{l}\neq t_{r}$ the working points given by $\varepsilon=\varepsilon_{M}$ (diagonal orange line seen in the in Fig.~\ref{fig:Plotlong}~(a) and see Appendix~\ref{ap:Comp}).

Comparing the resulting dephasing times considering noise coupled to the qubit through only one of the detuning parameters $\varepsilon$ or $\varepsilon_{M}$ (plotted in Fig.~\ref{fig:PlotIndi}~(a) and Fig.~\ref{fig:PlotIndi}~(e)), we find that the results are mirror symmetric to each other with the symmetry axis given by $\varepsilon=\varepsilon_{M}$. Single sweet spots are found on a straight vertical (horizontal) line passing through the center; serpentine vertical (horizontal) line for $t_{l}\neq t_{r}$ (a comparison of symmetric and asymmetric tunnel coupling can be found in Appendix~\ref{ap:Comp}).  

Considering transversal noise we cannot easily find an analytical expression for the $T_{\varphi}$ using the free decay model from Eq.~\eqref{eq:ramsey}. Thus, we have calculated $\delta\omega_{x}=|\delta\omega_{x,\varepsilon}|+|\delta\omega_{x,\varepsilon_{M}}|$ for $\delta\varepsilon=\delta\varepsilon_{M}\neq 0$ and $\delta t_{l}=\delta t_{r}=0$ which is a good measure for the coupling of noise to the qubit. In Fig.~\ref{fig:Plottrans}~(a) and (b) we plot the resulting $\delta\omega_{x}$ for different parameter settings. Note, that in these figures, $\delta\omega_{x}$ rather than the dephasing time is plotted, thus, small values lead to a longer lifetime of the qubit. Since transversal noise leads to transitions between the qubit states, this is also a first indication for the strength of the coupling between a qubit and a microwave cavity. Comparison of the results obtained for transversal charge noise (Fig.~\ref{fig:Plottrans} in Appendix~\ref{ap:trans}) and qubit-cavity coupling strength (Fig.~\ref{fig:plotQC}) show a high level of agreement as expected.

Inside the (1,1,1) charge configuration regime and considering only noisy detuning parameters $\varepsilon$ and $\varepsilon_{M}$ the sweet spot condition from Eq.~\eqref{eq:SScond} simplifies to
\begin{align}
	J\partial_{\varepsilon}J+3j\partial_{\varepsilon} j = J\partial_{\varepsilon_{M}}J+3j\partial_{\varepsilon_{M}} j=0
\end{align}
with $\partial_{\varepsilon,\varepsilon_{M}}J= \pm J_{l}^{2} (\varepsilon-\varepsilon_{M})/2t_{l}^{2}U + J_{r}^{2} (\varepsilon+\varepsilon_{M})/2t_{r}^{2}U$ and $\partial_{\varepsilon,\varepsilon_{M}}j= \pm J_{l}^{2} (\varepsilon-\varepsilon_{M})/2t_{l}^{2}U - J_{r}^{2} (\varepsilon-\varepsilon_{M})/2t_{r}^{2}U$. There exist only a single complete solution (DSS) in this regime for $\varepsilon=\varepsilon_{M}=0$\cite{Shim2016}. In contrast to the other 4 DSS the position of the center DSS is unaffected by the strength of the tunneling couplings $t_{l}$ and $t_{r}$, thus, more convenient for symmetric gate operations using the tunnel couplings as qubit control parameters.

\subsection{Tunnel noise}
\label{ssec:tunnel}

Symmetric qubit control by tuning the tunneling coupling between the QDs for qubit control has been proposed since the very beginning\cite{Loss1998} and recent experiments in Si/SiGe\cite{Reed2016} and GaAs\cite{Martins2016} indicate that symmetric operations lead to longer coherence times. Working at the symmetric operation points reduces the coupling to the charge noise originating from the detuning parameters, here, $\varepsilon$ and $\varepsilon_{M}$, hence, operating on a sweet spot relative to these parameters. However, this opens another channel for noise coupling to the qubit systems via fluctuations in the tunnel amplitude, since the tunneling is now gate-controlled, time-dependent, and strong, narrow-band filtering cannot be applied as effective as for the static case. Thus, the tunneling of the electrons is susceptible to charge noise.

In analogy to detuning noise, considering longitudinal noise through the tunnel parameters $t_{l}$ and $t_{r}$ can also drastically be reduced by working on sweet spots. Setting the noisy tunnel parameters $\delta t_{l}\neq 0$ and $\delta t_{r}\neq 0$ and ignoring noise coupled to the qubit through the detuning parameters $\delta\varepsilon=\delta\varepsilon_{M}=0$ we again find preferable working points and single sweet spots associated with either $t_{l}$ or $t_{r}$. The resulting dephasing is plotted in Fig.~\ref{fig:Plotlong}~(c)-(d) for the same parameter settings used previously. We find the best working points deep inside the (1,1,1) charge configuration regime, however, unlike in the case of detuning noise, there is no trace of DSS in the entire observed regime. The best working point we find is located at the center DSS ($\varepsilon=\varepsilon_{M}=0$) which is marginally better than the surrounding area while the other DSS at the charge transitions appear very unfavorable at first sight. A zoom-in, however, reveals a steep valley with a long dephasing time which is broadened by larger tunneling couplings. Therefore, the lifetime of the qubit at the DSS located at the charge transitions are limited by higher order effects. Strong tunnel coupling (see Fig.~\ref{fig:Plotlong}~(d)) drastically increases the lifetime at these points due to softening of the charge transitions challenging the center DSS. 

For the case $\delta t_{r}=0$ (see Fig.~\ref{fig:PlotIndi}~(e)-(f) in Appendix~\ref{ap:long}) we find single sweet spots nearby the charge transition associated with $t_{r}$, thus, $(1,1,1)\leftrightarrow(1,0,2)$ and $(1,1,1)\leftrightarrow(1,2,0)$ since at these lines in parameter space hopping from the left QD to the center QD is energetically highly unfavorable. The opposite case $\delta t_{l}=0$ is shown in  Fig.~\ref{fig:PlotIndi}~(g)-(h) in Appendix~\ref{ap:long}.

Inside the (1,1,1) charge configuration regime taking only noisy tunneling into account the sweet spot condition can be simplified to
\begin{align}
J_{l}(2J_{l}-J_{r})=J_{r}(2J_{r}-J_{l})=0.
\label{eq:tunSScond}
\end{align}
This condition is only fulfilled in the trivial case $J_{l}=J_{r}=0$, thus, there exist no DSS for tunneling noise. Single sweet spot corresponding to the tunneling parameter $t_{l}$ ($t_{r}$) require $J_{r,(l)}=2J_{l,(r)}$ which simplifies for $\varepsilon=0$ to $t_{r,(l)}=\pm\sqrt{2}t_{l,(r)}$. However, the best working points are given for an overall symmetric configuration including both tunneling and detuning. Since the DSS are all located at high symmetry points, the optimal working points are given by  $t_{l}=t_{r}$. Preferring points of operation nearby the states $\ket{3}$ and $\ket{4}$ ($\ket{2}$ and $\ket{5}$) the optimal ratio is $t_{l}/t_{r}>1$ ($t_{l}/t_{r}<1$). However, the benefit is not very large compared to operating on a DSS.

\subsection{Combination}
\label{ssec:comb}

Combining all effects, we plot in Fig.~\ref{fig:Plotlong}~(e)-(f) the dephasing time $T_{\varphi}$ taking into account all four noisy parameters $\varepsilon,\varepsilon_{M},t_{l},t_{r}$. Note that we put less weight to the tunneling parameters due to their smaller strength compared to the detuning parameters. As a result, we find that the previous areas with long coherence times considering only detuning noise of the sweet spots become less pronounced and softened due to the absence of DSS for tunneling. The center DSS still remains as the optimal point of operation in terms of pure coherence time, however, only slightly better than the surrounding area in the parameter space.

\section{Cavity quantum electrodynamics (c-QED)}
\label{sec:cqed}
\begin{figure}
\begin{center}
\includegraphics[width=1.0\columnwidth]{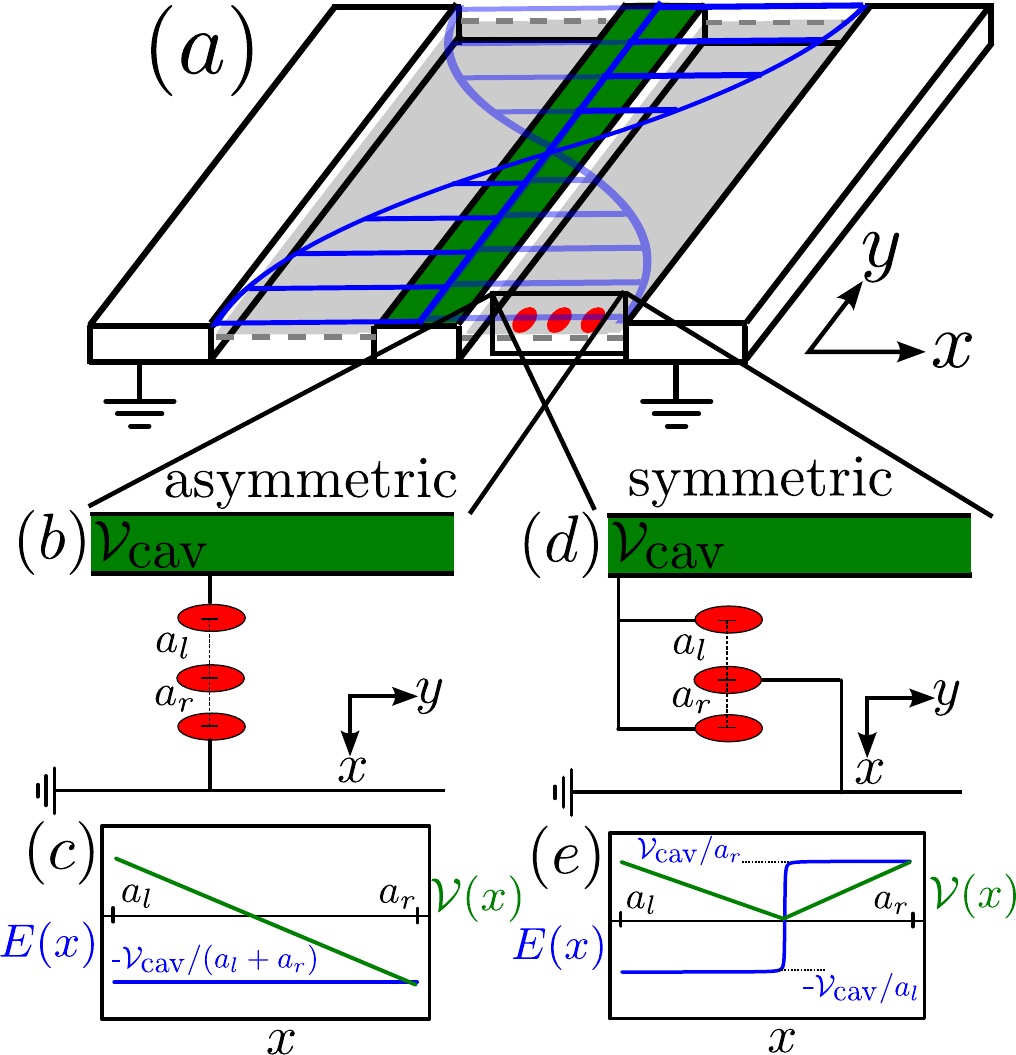}
\caption{$(a)$ Schematic illustration of a qubit implemented in a TQD coupled to the cavity and the architecture for a $(b)$ asymmetric and $(d)$ symmetric qubit-cavity coupling. The center conductor of the superconducting transmission line resonator is on the potential $\mathcal{V}\st{cav}$ while the outer conductors are connected to the ground to screen off surrounding fields. The corresponding potential (green) and electric field (blue) is shown for the asymmetric $(c)$ and symmetric $(e)$ arrangement as a function of the position $x$.}
\label{fig:arch}
\end{center}
\end{figure}
While the coupling to the uncontrolled fluctuations of the electric field at a three-spin qubit leads to dephasing, a controlled coupling to a quantized electromagnetic field in a microwave cavity can be used to couple qubits over long distances.
We consider three-spin qubits realized in a linear TQD embedded in a superconducting transmission line resonator with a single photon mode near the resonance frequency of the qubit.  Analogous to Sec.~\ref{Sec:chargedephasing}, we calculate the qubit-cavity coupling for the full ($\varepsilon$,$\varepsilon_{M}$)-plane including all charge configuration numerically, and subsequently, we approximate the center of the (1,1,1) charge configuration analytically in order to analyze the results. To generalize our previous analysis\cite{Russ2015b} to the full range of charge states studied in the previous sections, we extend the existing formalism to include all six relevant states given by Eqs.~\eqref{state:0}-\eqref{eq:logicalspace}. We model the dipolar interaction\cite{Cohen1989} between the qubit and the cavity with
\begin{align}
H\st{QC}&=-e\boldsymbol{E} \cdot \boldsymbol{\hat{x}}=-e\boldsymbol{\mathcal{E}}\cdot \boldsymbol{\hat{x}}\,(a+a^{\dagger})
\label{eq:GenHam}
\end{align}
and define the qubit-cavity coupling strength as
\begin{align}
	g\equiv -e\bra{0}\boldsymbol{\mathcal{E}} \cdot \boldsymbol{\hat{x}}\ket{1},
\label{eq:GenCoup}
\end{align}
where $a^{\dagger}$ ($a$) creates (annihilates) a photon with frequency $\omega_{ph}$ of the cavity mode.
Note, that in this paper the formalism using $\boldsymbol{E} \cdot \boldsymbol{\hat{x}}$ is more convenient than the equivalent formalism\cite{Cohen1989} $\boldsymbol{A} \cdot \boldsymbol{\hat{p}}$ used in previous works\cite{Burkard2006,Russ2015b}. Here, $\boldsymbol{E}$ is the quantized electric field, $\boldsymbol{E}=\boldsymbol{\mathcal{E}}(a+a^{\dagger})$, and $\boldsymbol{A}$ is the quantized electromagnetic vector potential. 

In Fig.~\ref{fig:arch}~$(a)$ the basic implementation is schematically shown together with the two architectures discussed in this work which we label asymmetric and symmetric coupling corresponding to the affected detuning parameter. In this setting, the qubit is placed in the anti-node of the electromagnetic field of the cavity to achieve the strongest coupling. The vacuum coupling strength of the interaction is $g_{0}$, here, defined as the coupling strength of the qubit if the dipole moment $\bra{0}\hat{x}\ket{1}\cong a_{l}+a_{r}$, thus, the full length of the TQD, where $a_{l,(r)}$ is the distance between the left (right) QD and the center QD (sketched in Fig.~\ref{fig:arch}~(b) and (d)). We find
\begin{align}
	g_{0}=-e\mathcal{E}(a_{l}+a_{r})
\end{align}
with $\mathcal{E}=|\boldsymbol{\mathcal{E}}|=\sqrt{\hbar\omega\st{ph}/2\epsilon_{0}\epsilon v}$, where $e$ is the electron charge, $\epsilon_{0}$ ($\epsilon$) is the (relative) dielectric constant of the vacuum (material) and $v$ is the volume of the cavity\cite{Cohen1989}. Using realistic parameter settings for a silicon TQD ($v=\unit[3]{cm}\times\unit[5]{\mu m}\times\unit[5]{\mu m}$, $\epsilon\approx 12$, $\omega\st{ph}=\unit[4.7]{GHz}$, and $a_{l}+a_{r}=\unit[60]{nm}$) we obtain $g_{0}=\unit[2\pi\times1.96]{MHz}$. Note, that this is the pure vacuum coupling strength and no field enhancement was included, which can further enhance the strength drastically\cite{Guilherme2014,Samkharadze2016}. To make connection with experiments, it is sometimes more convenient to express the vacuum coupling strength in terms of capacitance and impedance, thus, $\mathcal{E}=\nu \omega\st{ph}\sqrt{Z_{0}/\pi\hbar}/2w$ with $\nu=C\st{con}/(C\st{con}+C\st{TQD})$. Here, $\nu$ is the relative capacitance of the TQD, $C\st{TQD}$, and the capacitance of the connection to the resonator, $C\st{Con}$, $Z_{0}$ is the characteristic impedance and $w$ the distance in which the voltage drop occurs\cite{Srinivasa2016}. Recent experiments show high impedance resonators giving rise to a vacuum coupling strength $g_{0}$ in the order of $\unit[2\pi\times100]{MHz}$\cite{Samkharadze2016}.

We first construct the real-space wave-functions of the states $\ket{0},\ket{1},\ket{2},\ket{3},\ket{4}$, and $\ket{5}$, needed for the transition dipole matrix element. For this we use the formalism of orthonormalized Wannier orbitals\cite{Russ2015b}, which transforms overlapping single-electron wave-functions $\ket{\phi_{i}}$ into an orthonormal basis of maximally localized\cite{Marzari2012} wave-functions $\ket{\Phi_{i}}$ with $i\in\lbrace 1,2,3\rbrace$. Here, the overlaps between the pure single-electron wave-functions are denoted as $S_{l}\equiv\braketv{\phi_{1}}{\phi_{2}}$, $S_{r}\equiv\braketv{\phi_{2}}{\phi_{3}}$, and $0=\braketv{\phi_{1}}{\phi_{3}}$ due to the linear arrangement\cite{Russ2015b}. As a result we obtain for the position operator in the basis $\left\lbrace \ket{0},\ket{1},\ket{2},\ket{3},\ket{4},\ket{5}\right\rbrace$
\begin{align}
\hat{x}=\left(
	\begin{smallmatrix}
	 0 & 0 & \frac{1}{\sqrt{2}}x_{12} & \frac{1}{\sqrt{2}}x_{32} & \frac{1}{\sqrt{2}}x_{23}  & \frac{1}{\sqrt{2}}x_{21}  \\
	 0 & 0 & \sqrt{\frac{3}{2}} x_{12} & -\sqrt{\frac{3}{2}} x_{32} & -\sqrt{\frac{3}{2}} x_{23} & \sqrt{\frac{3}{2}} x_{21} \\
	 \frac{1}{\sqrt{2}}x_{21} & \sqrt{\frac{3}{2}} x_{21} & x_{11}-x_{22} & -x_{31} & 0 & 0 \\
	 \frac{1}{\sqrt{2}}x_{23} & -\sqrt{\frac{3}{2}} x_{23} & -x_{13} & x_{33}- x_{22} & 0 & 0\\
	 \frac{1}{\sqrt{2}}x_{32} & -\sqrt{\frac{3}{2}} x_{32} & 0 & 0  & x_{22}-x_{33} & x_{31} \\
	 \frac{1}{\sqrt{2}}x_{12} & \sqrt{\frac{3}{2}} x_{12}  & 0 & 0 & x_{13} & x_{22}- x_{11}\\
	\end{smallmatrix}
	\right),
	\label{eq:positionoperator}
\end{align}
where $x_{ij}\equiv\bra{\Phi_{i}}\hat{x}\ket{\Phi_{j}}$ denotes the transition dipole matrix element between the Wannier orbitals. Under the assumption of equal confinement potentials in each QD these transition dipole matrix elements can always be chosen real\cite{Russ2015b}.

\subsection{Asymmetric architecture}

\begin{figure*}
\begin{center}
\includegraphics[width=1.\textwidth]{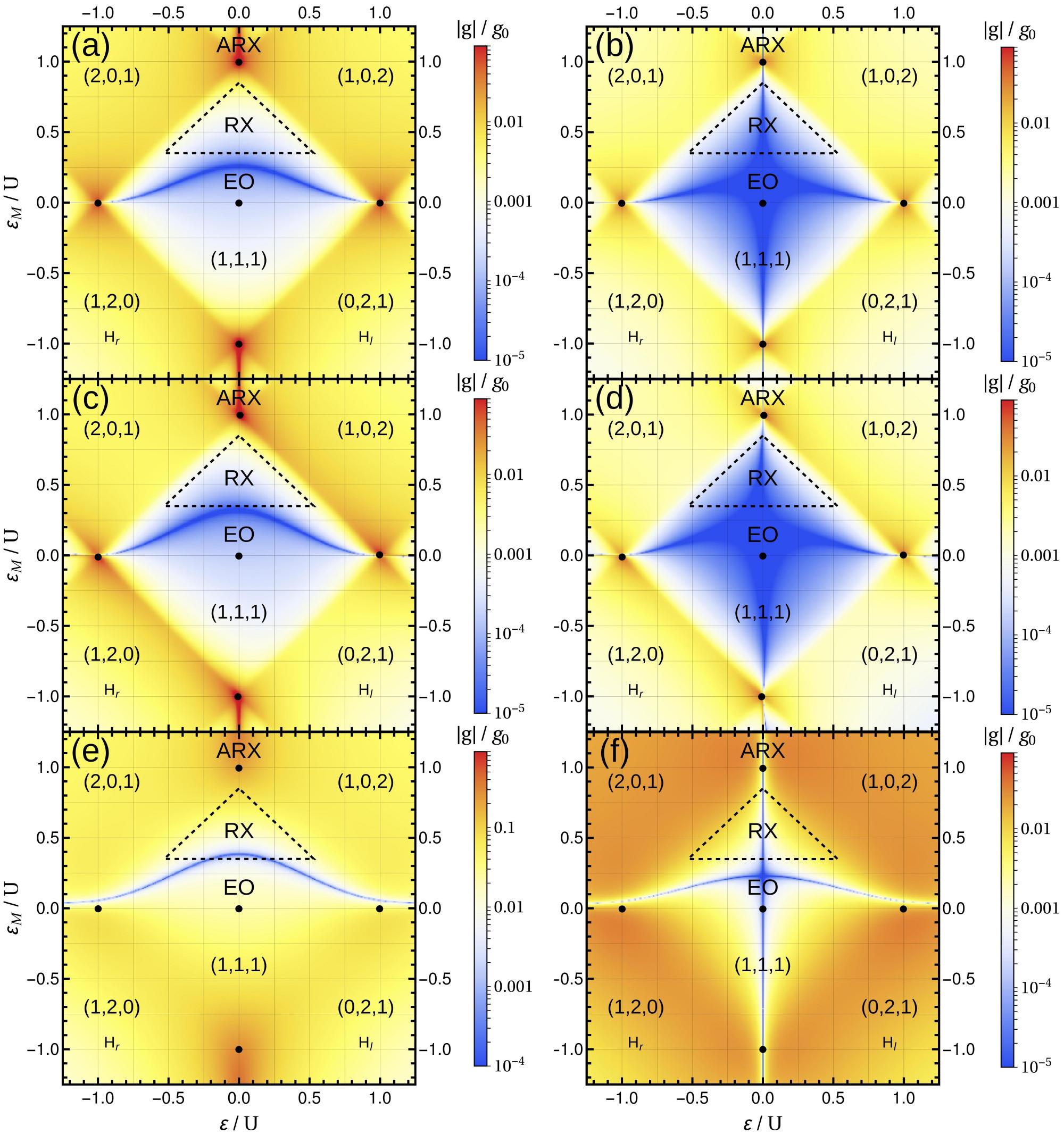}
\caption{The qubit-cavity coupling strength $g$ in units of the vacuum coupling strength $g_{0}$ as a function of the detuning parameters $\varepsilon$ and $\varepsilon_{M}$ for the asymmetric coupling (left column) and symmetric coupling (right column). The parameters are chosen as follows; top row ((a) and (b)) $t_{l}=t_{r}=\unit[20]{\mu eV}$, center row ((c) and (d))  $t_{l}=\unit[22]{\mu eV}$ and $t_{r}=\unit[15]{\mu eV}$, and bottom row ((e) and (f)) $t_{l}=t_{r}=\unit[200]{\mu eV}$. The interdot distances $a_{l}$ and $a_{r}$ and the overlaps $S_{l}$ and $S_{r}$ are set by the strength of the tunneling parameters\cite{Burkard1999,Russ2015b}. The black dots denote the DSS.}\label{fig:plotQC}
\end{center}
\end{figure*}

Placing the TQD inside the cavity such that the electric field aligns with the long axis of the qubit\cite{Russ2015b} (see Fig.~\ref{fig:arch}~(b)), leads to  a standard dipole coupling interaction between the qubit and the cavity. States with an asymmetric charge configuration interact with the electromagnetic field of the cavity via their coupling to the opposite charge state, e.g., $\ket{2}\leftrightarrow\ket{3}$ and $\ket{4}\leftrightarrow\ket{5}$, while creating or annihilating a photon in the process. Hence, the qubit-cavity interaction $g_{A}$ in Eq.~\eqref{eq:GenHam} can be simplified to
\begin{align}
H_{A}=-e\mathcal{E} \,\, \hat{x}(a+a^{\dagger}),
\label{eq:AIHam}
\end{align}
and the qubit-cavity coupling strength for the asymmetric architecture becomes 
\begin{align}
g_{A}\equiv-e\mathcal{E} \bra{g} \hat{x}\ket{e},
\label{eq:coupSym}
\end{align}
where $\ket{g}$ is the ground and $\ket{e}$ the first excited state. For a (1,1,1) charge configuration these states coincide with $\ket{0}$ and $\ket{1}$. In Fig.~\ref{fig:plotQC} (left column) the qubit-cavity coupling is calculated numerically and plotted for various parameter settings. The weakest qubit-cavity coupling can be found inside the (1,1,1) charge configuration regime which is expected due to the symmetric electron distribution. The strongest coupling is located near the four outer DSS, since at these points a charge transfer only requires a small variation of the detuning parameters to produce a large dipole moment. The asymmetric implementation favors a charge transition associated with a transfer of one electron from the left QD to the right QD\cite{Russ2015b}, thus, $\ket{2}\leftrightarrow\ket{3}$ and $\ket{3}\leftrightarrow\ket{4}$ resulting in a strong coupling along a vertical line above and underneath the (1,1,1) charge regime in Fig.~\ref{fig:plotQC}.

Deep inside the (1,1,1) charge configuration regime the ground states are $\ket{\tilde{0}}$ and $\ket{\tilde{1}}$ which are $\ket{0}$ and $\ket{1}$ hybridized by a small admixture of the other charge states ($\ket{2}$, $\ket{3}$, $\ket{4}$, $\ket{5}$), hence, $\tilde{x}=\E^{S}\hat{x}\E^{-S}\approx \hat{x} + \left[S,\hat{x}\right]+\frac{1}{2}\left[S,\left[S,\hat{x}\right]\right]+\cdots$. Here, $S$ is the same Schrieffer-Wolff transformation matrix used to derive the qubit splitting in the low-energy subspace given in Eq.~\eqref{eq:HeigenSW}. As a result, we obtain a closed analytical expression for the qubit-cavity coupling strength in Eq.~\eqref{eq:coupSym},
\begin{align}
\begin{split}
	g_{A}/g_{0} =
	& -\sqrt{\frac{3}{2}}\left[ \frac{J_{l}}{t_{l}}\frac{\text{Re}(x_{12})}{2(a_{l}+a_{r})}-\frac{J_{r}}{t_{r}}\frac{\text{Re}(x_{23})}{2(a_{l}+a_{r})}\right]\\
		- \frac{\sqrt{3}}{4}\bigg[& \frac{\left(\varepsilon-\varepsilon_{M}\right)}{U}  \frac{J_{l}^{2}}{t_{l}^{2}}\frac{x_{11}}{a_{l}+a_{r}} +\frac{\left(\varepsilon+\varepsilon_{M}\right)}{U}  \frac{J_{r}^{2}}{t_{r}^{2}} \frac{x_{33}}{a_{l}+a_{r}}\bigg].
		\end{split}
		\label{eq:gSW}
\end{align}
Here, the first (second) term in the second line resembles the matrix element of a DQD in the left (right) QD and the center QD\cite{Burkard2006} which compensate each other at the EO DSS located at $\varepsilon=0$ and $\varepsilon_{M}=0$. Due to the sign change in the first term the overall matrix element is nonzero at the EO DSS as may be expected from general considerations. For a completely symmetric setup, $\varepsilon=0$, $a_{l}=a_{r}\equiv a$, $S_{l}=S_{r}\equiv S_{0}$, and $t_{l}=t_{r}\equiv t$ which leads to $J_{l}=J_{r}\equiv J_{0}=2t^{2}U/(U^{2}-\varepsilon_{M}^{2})$, $\text{Re}(x_{12})=-\text{Re}(x_{23})\approx-3aS_{0}$, and $x_{11}=-x_{33}\approx-a$, Eq.~\eqref{eq:gSW} simplifies to
\begin{align}
g_{A,0}/g_{0} &=-\frac{\sqrt{3}}{2}\,\frac{\,\varepsilon_{M}\,J_{0}}{U^{2}-\varepsilon_{M}^{2}} +\sqrt{\frac{3}{2}}\, \frac{3\,J_{0}\,S_{0}}{2t}.
\end{align}
The first term is identical to the expression for a simple charge model $g_{A,0}/g_{0} =-\sqrt{(\partial_{\varepsilon}J)^{2}+3(\partial_{\varepsilon}j)^{2}}/2$ for this choice of parameters\cite{Srinivasa2016} and approaches zero at the DSS while the second term remains finite. The general expression, however, is given in Eq.~\eqref{eq:gSW}. Introducing $\xi\equiv S_{0}/t$, we find zero qubit-cavity coupling at $\xi=\sqrt{2/3}\,\varepsilon_{M}/(U^{2}-\varepsilon_{M}^{2})$, e.g., for the exchange-only DSS $\varepsilon_{M}=0$ the condition is $\xi=0$, thus, $S_{0}\ll t$.

\subsection{Symmetric architecture}

Alternatively, one can place the TQD in the cavity such that the center QD is connected to the transmission line while the outer two QDs are connected with the ground plane\cite{Russ2015b} (see Fig.~\ref{fig:arch}~(d)). In this scheme, the electric field is not alined continuously with the $x$-axis or other axis, but rather, it changes sign and strength in the center. Fig.~\ref{fig:arch}~(e) shows the expected electric field as a function of position which without screening effects can be described as a jump function. To model the electric field, we use
\begin{align}
\mathcal{E}(\hat{x})&=\frac{1}{\pi}\left\lbrace\tan^{-1}\left[\frac{\hat{x}}{T\left(a_{l}+a_{r}\right)}\right]+\frac{\pi(a_{l}-a_{r})}{2(a_{l}+a_{r})}\right\rbrace,
\end{align}
where $T$ is a screening parameter that softens the step (see Fig.~\ref{fig:arch}~(e)). Note that $\mathcal{E}$ is an operator here because it is a function of the position operator, hence, we obtain for the qubit-cavity interaction
\begin{align}
H_{S}&=-e\,\mathcal{E}(\hat{x})\,\hat{x}\,(a+a^{\dagger}).
		%&=-\frac{E_0}{ \pi } \tan^{-1}\left[\frac{\left(a_{l}+a_{r}\right)\hat{x}}{T}\right]+\frac{a_{l}-a_{r}}{2(a_{l}+a_{r})}
\label{eq:SIHam}
\end{align}
This Hamiltonian can be understood as a generalization of the single-mode dipole interaction\cite{Cohen1989} $H\st{dip}=-e\boldsymbol{E}\cdot\boldsymbol{x}$, in which the electric field $\boldsymbol{E}(\boldsymbol{\hat{x}})$ can be dependent of the position $\boldsymbol{\hat{x}}$ associated with the architecture of the qubit inside the cavity. 
For the qubit-cavity strength for the symmetric architecture we find
\begin{align}
g_{S}\equiv -e\bra{g} \mathcal{E}(\hat{x})\,\hat{x}\ket{e},
\label{eq:coupAsym}
\end{align} 
with $\ket{g}$ again being the ground state and $\ket{e}$ the first excited state. Unfortunately, $g_{S}$ cannot be expressed in a closed analytical form in the general case. In Fig.~\ref{fig:plotQC} (right column), the results are numerically calculated and plotted for the same parameter settings as for the asymmetric architecture (left column). We find the weakest values for the qubit cavity coupling again deep inside the (1,1,1) charge configuration regime, which is expected due to the large energy required to enable a charge transition. In the vicinity of the expected charge transition areas which includes the DSS we find the strongest coupling strength. For $\varepsilon=0$ and $t_{l}=t_{r}$ (for $t_{l}\neq t_{r}$ slightly shifted) the symmetric coupling $g_{S}$ vanishes since for this architecture a charge transition between (1,0,2) and (2,0,1) or (1,2,0) and (0,2,1) is unfavorable with both outer QDs being at the same potential. In contrast to the asymmetric architecture, the symmetric implementation should favor a charge transition associated with an electron transfer only between the left (right) and center QD, thus, $\ket{2}\leftrightarrow\ket{4}$ ($\ket{3}\leftrightarrow\ket{5}$). Thus, we expect a strong response seen in a horizontal line from left to right in the $(\varepsilon,\varepsilon_{M})$ parameter plane through the center (see Fig.~\ref{fig:plotQC}). We believe the reason for the absence of this line in the numerical results (Fig.~\ref{fig:plotQC}, right column) is the need of two simultaneous charge transfers, hence, a two photon process which is beyond the scope of this model.

Inside the (1,1,1) charge configuration regime and assuming a large screening $T> 1$, thus, justifying an expansion of the position dependent electric field, $\left.E(\hat{x})/\mathcal{E} \approx \hat{x}/\pi T (a_{l}+a_{r})+ (a_{l}-a_{r})/2(a_{l}+a_{r})\right.$ we find an analytical expression for the qubit-cavity coupling defined in Eq.~\eqref{eq:coupAsym}
\begin{align}
g_{S}/g_{0}=  \frac{a_{l}-a_{r}}{2(a_{l}+a_{r})^{2}}\bra{g}\hat{x}\ket{e}+\frac{1}{\pi T(a_{l}+a_{r})^{2}}\bra{g}\hat{x}^{2}\ket{e}
\label{eq:gsSW}
\end{align} 
The first term $\bra{g}\hat{x}\ket{e}$ is the asymmetric coupling given in Eq.~\eqref{eq:gSW} and for the second term we obtain analogously 
\begin{align}
\begin{split}
	\bra{g} \hat{x}^{2}\ket{e} =
	& \frac{\sqrt{3}}{4}\left( \frac{J_{l}^{2}}{t_{l}^{2}}\frac{x_{11}^{2}}{2}-\frac{J_{r}^{2}}{t_{r}^{2}}\frac{x_{33}^{2}}{2}\right)\\
	&+\frac{\sqrt{3}}{4}\left[ \text{Re}(x_{12}) +
		\frac{(\varepsilon-\varepsilon_{M})}{U}  \frac{J_{l}}{t_{l}}\frac{x_{11}}{\sqrt{2}}\right]^{2} \\
		&-\frac{\sqrt{3}}{4}\left[ \text{Re}(x_{23}) -
		\frac{(\varepsilon+\varepsilon_{M})}{U}  \frac{J_{r}}{t_{r}}\frac{x_{33}}{\sqrt{2}}\right]^{2} .
				\end{split}
				\label{eq:xsSW}
\end{align}
For a completely symmetric setup, $\varepsilon=0$, $a_{l}=a_{r}\equiv a$, $S_{l}=S_{r}\equiv S_{0}$, and $t_{l}=t_{r}\equiv t$ which leads to $J_{l}=J_{r}\equiv J_{0}$, $\text{Re}(x_{12})=-\text{Re}(x_{23})\approx-3aS_{0}$ and $x_{11}=-x_{33}\approx-a$. Thus, in this fully symmetric case, both Eq.~\eqref{eq:xsSW} and Eq.~\eqref{eq:gsSW} yield $g_{A}=0$.
This result is consistent with previous results using a simple phenomenological approach\cite{Russ2015b}.

\section{Conclusion}
\label{sec:conc}

In this work we have analyzed different types of three-spin-$\frac{1}{2}$ qubits in an electric environment, either coupled to charge noise or to coherent electric fields in a superconducting strip-line cavity. The first coupling needs to be minimized or eliminated in order to achieve long-lived qubits. On the other hand we want to maximize and control the coupling between the qubit and the electric field of the cavity in order to acquire the strong coupling regime needed for a fast long-distance two-qubit gate\cite{Russ2015b}.

In the case of a fluctuating electromagnetic environment we have provided an extended description considering external electric fluctuations coupled to the qubit through four distinct noisy parameters, two detuning parameters, and two tunneling parameters. We presented and discussed the best suitable working points which take all these couplings into account and minimize the impact limiting the coherence time at the detuning sweet spot. However, no quadruple sweet spot was found suppressing first order noise effects of all four noisy parameters simultaneously. We found that charge noise coupled to the tunneling parameters is the limiting factor due to the possibility of working on one of the five double sweet spots (DSS) for noisy detuning parameters. Four of the five DSS are located each at the crossover regions between connecting asymmetric charge configurations with the fifth sitting in the center of the (1,1,1) charge configuration regime. We have presented a full map of the dephasing time in the $(\varepsilon,\varepsilon_{M})$ parameter plane taking either the effect of all four noisy parameters, pairs of two noisy parameters or each noisy parameter individually into account. The optimal strategy depends on the strength of the noise and the strength of each tunneling parameter, however, it appears that a symmetric implementation ($t_{l}=t_{r}$) typically provides the best result exactly at the DSS while a slightly asymmetric implementation ($t_{l}\neq t_{r}$) elongates the favorable area surrounding the DSS. 

In the second part of the paper, we have presented a full description of the coupling between the qubit and a high-finesse transmission line cavity taking both basic alignments of connecting the physical qubit and the cavity into account, an asymmetric one being the intuitive where the first and the last QDs are on opposite potentials with a constant electric field. For the symmetric coupling both outer QDs are on the same potential while the center QD is connected with the transmission line of the cavity. For both alignments we have provided a detailed map of the coupling strength in parameter space and derived analytical results inside the (1,1,1) charge regime fully agreeing and extending previous results where they exist. Additional features only appearing in the extended model were discussed. Best working points for the asymmetric alignment were located nearby the $(2,0,1)\leftrightarrow (1,0,2)$ and $(1,2,0)\leftrightarrow (0,2,1)$ charge transitions (the exact position depending on the parameter setting) featuring the top and bottom DSS as favorable choices. For the symmetric alignment these points turn out to be less favorable within the scope of our model and working points nearby the (1,1,1) charge transitions should be favored in order to obtain decent coupling strength combined with long coherence of the qubit. However, for the symmetric architecture  we expect additional influence of two-photon processes which are beyond the scope of this paper.

\acknowledgments

We acknowledge funding from ARO through Grant No. W911NF-15-1-0149.

\appendix

\begin{figure*}
\begin{center}
\includegraphics[height=0.85\textheight]{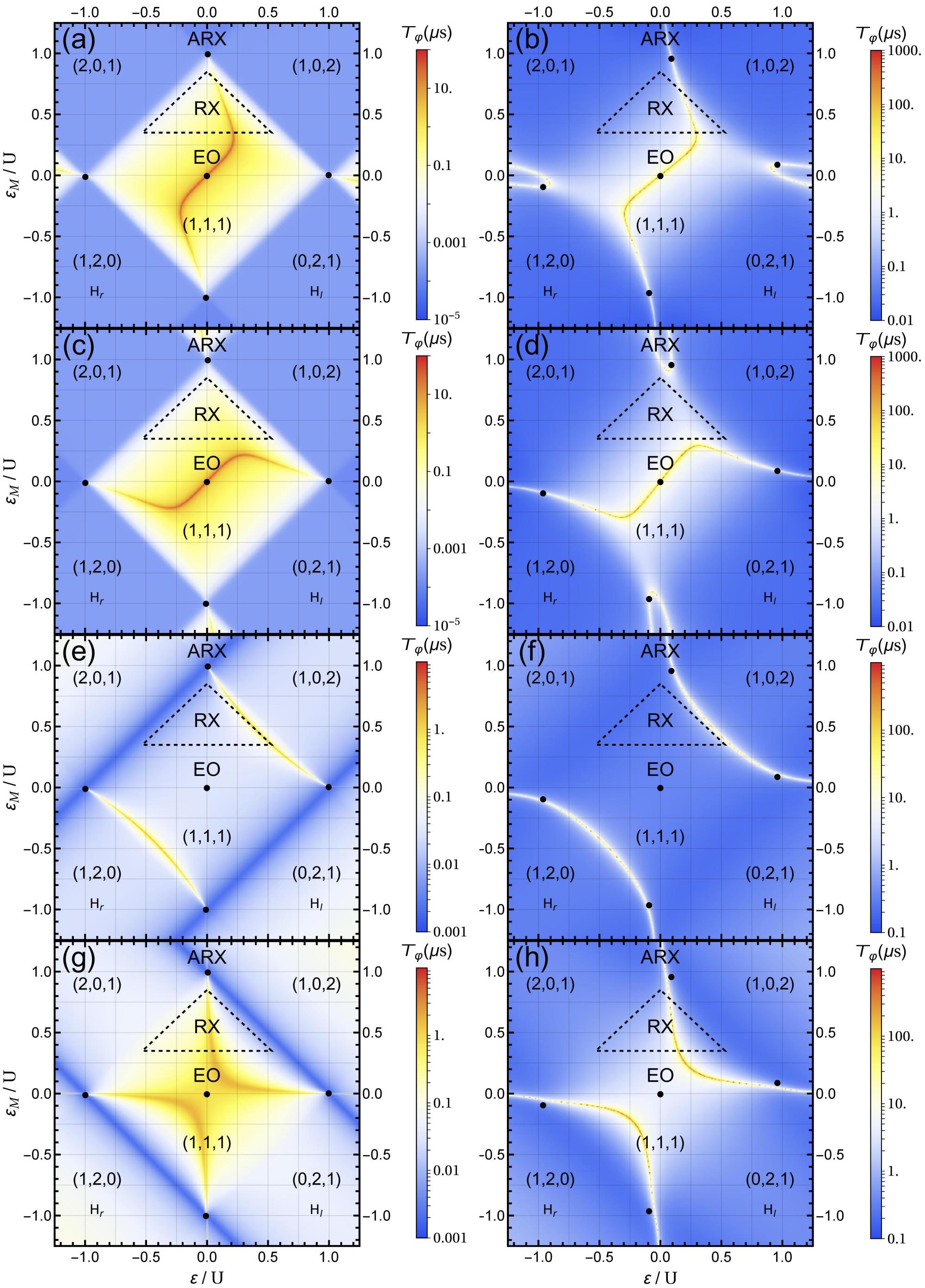}
\caption{Dephasing time $T_{\varphi}$ due to longitudinal noise for each noisy parameter $\varepsilon,\,\varepsilon_{M},\,t_{l}$, and $t_{r}$ individually in the $(\varepsilon,\varepsilon_{M})$-plane. Each row shows the dephasing time due to a single noisy parameter (from top to bottom: $\varepsilon,\,\varepsilon_{M},\,t_{l},\,t_{r}$), while we choose the parameter settings identical in each column. The left column contains the results for weak tunneling and strong noise, while the right column comprises the results for strong tunneling and weak noise. Parameters are set as follows; $t_{l}=\unit[22]{\mu eV}$, $t_{r}=\unit[15]{\mu eV}$, $A_{q}=\unit[1]{\mu eV}^{2}$ where $q=\varepsilon$ in (a) and $q=\varepsilon_{M}$ in (c), and $A_{q}= (\unit[10^{-1}]{\mu eV})^{2}$, where $q=t_{l}$ in (e) and $q=t_{r}$ in (g), for the left column and $t_{l}=\unit[220]{\mu eV}$, $t_{r}=\unit[150]{\mu eV}$, $A_{q}=(\unit[10^{-2}]{\mu eV})^{2}$ where $q=\varepsilon$ in (b) and $q=\varepsilon_{M}$ in (d), and $A_{q}= (\unit[10^{-3}]{\mu eV})^{2}$, where $q=t_{l}$ in (f) and $q=t_{r}$ in (h), for the right column. The black dots represent DSS.}
\label{fig:PlotIndi}
\end{center}
\end{figure*}

\begin{figure*}
\begin{center}
\includegraphics[width=1.\textwidth]{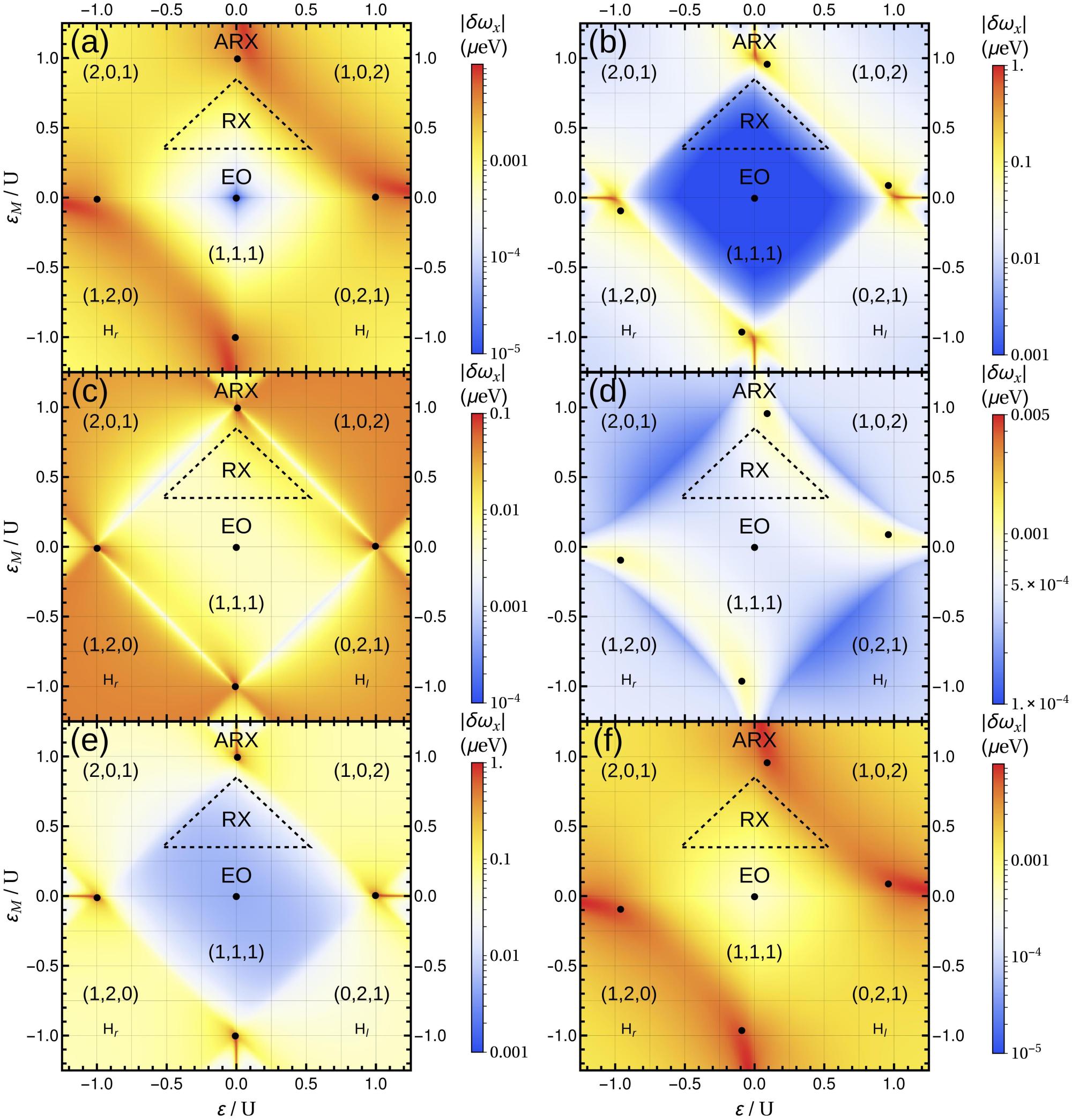}
\caption{Impact of transversal noise as a function of the detuning parameters $\varepsilon$ and $\varepsilon_{M}$. In this figure $\delta\omega_{x}$ rather than the dephasing time is plotted, thus, small values lead to longer coherence times of the qubit. In the top row ((a) and (b)) we consider charge noise only from the two detuning parameters $\varepsilon$ and $\varepsilon_{M}$, in the center row ((c) and (d)) we consider charge noise only from the two tunneling parameters $t_{l}$ and $t_{r}$, and in the bottom row ((e) and (f)) we consider charge noise from all four parameters simultaneously, while we choose the parameter settings identical in each column. The left column contains the results for weak tunneling and strong noise, while the right column comprises the results for strong tunneling and weak noise. The black dots indicate DSS.
Parameters are set as follows; $t_{l}=\unit[22]{\mu eV}$, $t_{r}=\unit[15]{\mu eV}$, $\delta q=\unit[1]{\mu eV}$ where $q=\varepsilon,\varepsilon_{M}$ in (a) and (e), and $\delta q= \unit[10^{-1}]{\mu eV}$, where $q=t_{l},t_{r}$ in (c) and (e), for the left column and $t_{l}=\unit[220]{\mu eV}$, $t_{r}=\unit[150]{\mu eV}$, $\delta q=\unit[10^{-2}]{\mu eV}$ where $q=\varepsilon,\varepsilon_{M}$ in (b) and (f), and $\delta q= \unit[10^{-3}]{\mu eV}$, where $q=t_{l},t_{r}$ in (d) and (f), for the right column.}
\label{fig:Plottrans}
\end{center}
\end{figure*}

\begin{figure*}
\begin{center}
\includegraphics[width=1.\textwidth]{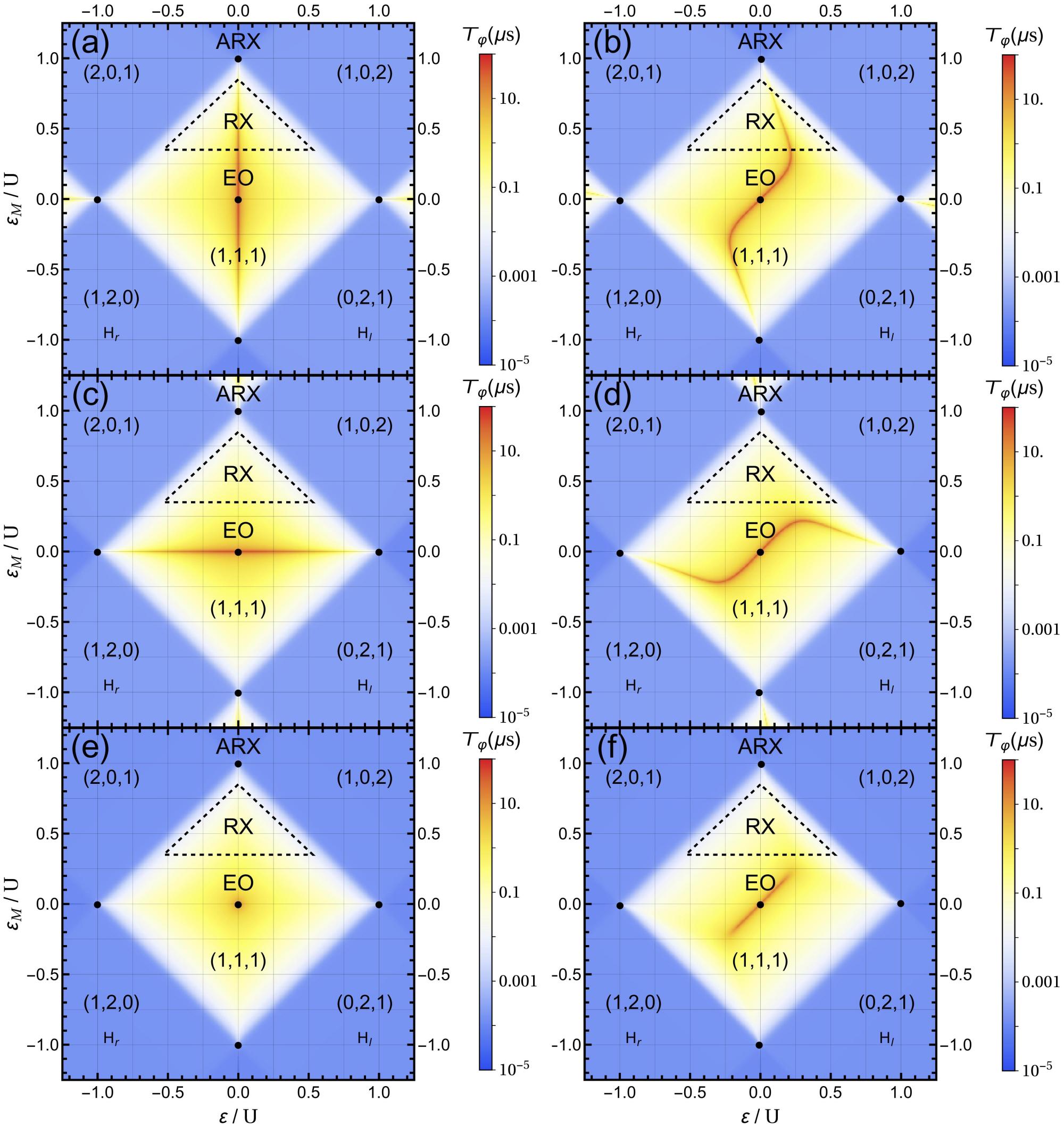}
\caption{Comparison of the dephasing time $T_{\varphi}$ as a function of the detuning parameters $\varepsilon$ and $\varepsilon_{M}$ for symmetric and asymmetric tunnel coupling. In the top row ((a) and (b)) we consider charge noise only from $\varepsilon$, in the center row ((c) and (d)) we consider charge noise only from $\varepsilon_{M}$, and in the bottom row ((e) and (f)) we consider charge noise both detuning parameters, $\varepsilon$ and $\varepsilon_{M}$, simultaneously, while we choose the parameter settings identical in each column. The left column comprises the results for symmetric tunneling, while the right column repeats the results for asymmetric tunneling given in Fig.~\ref{fig:Plotlong}~(a) and Fig.~\ref{fig:PlotIndi}~(a) and (c). The black dots denote DSS. Parameters are set as follows; $t_{l}=t_{r}=t=\unit[20]{\mu eV}$ for the left column and $t_{l}=\unit[22]{\mu eV}$, $t_{r}=\unit[15]{\mu eV}$ for the right column. Further we set $A_{q}=\unit[1]{\mu eV}^{2}$ where $q=\varepsilon$ in (a) and (b), $q=\varepsilon_{M}$ in (c) and (d), and $q=\varepsilon,\varepsilon_{M}$ in (e) and (f).}
\label{fig:Comp}
\end{center}
\end{figure*}

\section{Longitudinal noise coupled only through a single noisy parameter}
\label{ap:long}

In Fig.~\ref{fig:PlotIndi} the dephasing times are plotted considering the case where charge noise is coupled to the qubit through only a single noisy parameter for two different parameter settings. Comparing the plots, we find that the resulting dephasing times for detuning noise in $\varepsilon$ and $\varepsilon_{M}$ are mirror symmetric to each other with the symmetry axis given by $\varepsilon=\varepsilon_{M}$, while there is no such symmetry axis for tunneling noise in the general case $t_{l}\neq t_{r}$. The SSS for detuning noise are located on a serpentine vertical or horizontal line with the crossing points given by the DSS. For tunneling noise ($t_{l}$ or $t_{r}$) we find that the SSS are located on a narrow curve connecting the top DSS and the right DSS as well as the right DSS and the bottom DSS. A zoom in, however, reveals that the DSS are not directly located on the line, more precisely, the SSS for tunneling noise in $t_{l}$ ($t_{r}$) are slightly shifted to the right (left) in parameter space. In contrast to detuning noise, there exist no crossing points for the SSS in tunneling noise.

\section{Transversal noise}
\label{ap:trans}

In Fig.~\ref{fig:Plottrans} we plot the transversal effect of charge noise for the same parameter settings as for longitudinal noise. Since the dephasing time cannot be calculated easily for transversal noise, we plot instead $\delta\omega_{x}=\sum_{q\neq p}|\delta\omega_{x,q}|+|\delta\omega_{x,p}|$ with $q,p\in\lbrace\varepsilon,\varepsilon_{M},t_{l},t_{r}\rbrace$ and $\delta\omega_{x,q}$ given in Sec.~\ref{ssec:genframe} which is a good measure for the susceptibility of the noise. 

In Fig.~\ref{fig:Plottrans} the combined effects of only noisy detuning (top row), only noisy tunnel coupling (center row), and the combined effects of two noisy detuning and tunneling parameters (bottom row) are shown. Comparison leads to similar results as for longitudinal noise. The well protected spot in the center of the (1,1,1) charge configuration considering only detuning noise fades away in the bottom row due to the influence of tunneling noise and becomes as well protected against transversal charge noise as its surrounding are. Since the outer DSS are located at charge transitions, they are very susceptible to transversal noise and, therefore, will dephase much faster than other points. A zoom-in, however, reveals that the outer DSS are not located at a maximum, although, much more susceptible than the center DSS to transversal charge noise.

\section{Dephasing for symmetric tunnel coupling}
\label{ap:Comp}

In Fig.~\ref{fig:Comp} we plot the resulting dephasing time $T_{\varphi}$ for symmetric tunnel couplings $t_{l}=t_{r}=t$ (left column) and asymmetric tunnel coupling $t_{l}\neq t_{r}$ (right column) taking into account only the noise from the detuning parameters. Comparing these two situations, we find that in the case of only a single noisy parameter $\varepsilon$ ($\varepsilon_{M}$) and symmetric tunneling the SSS can be found on a straight vertical (horizontal) line through the center DSS in contrast to the serpentine vertical (horizontal) lines for asymmetric tunneling. Taking both noisy parameters into account, this leads to a single crossing point of the two lines at the center DSS in the symmetric case ($t_{l}=t_{r}=t$) while we find an elongated area in the asymmetric case ($t_{l}\neq t_{r}$) for the center DSS. Therefore, the asymmetric case allows for a greater flexibility in choosing the point of operation while still being protected against longitudinal charge noise.

\bibliography{lit4}

\end{document}